\documentclass[prd,superscriptaddress,twocolumn,showpacs,nofootinbib]{revtex4}

\usepackage{amsfonts}
\usepackage{amsmath}
\usepackage{amssymb}
\usepackage{bm}
\usepackage{dcolumn}
\usepackage{epsfig}
\usepackage{graphicx}
\usepackage{graphics}
\usepackage[latin1]{inputenc}
\usepackage{latexsym}
\usepackage{rotating}
\usepackage{hyperref}
\usepackage[usenames]{color}

\newcommand\be{\begin{equation}}
\newcommand\ba{\begin{eqnarray}}
\newcommand\ee{\end{equation}}
\newcommand\ea{\end{eqnarray}}

\newcommand{\BH}{{\mbox{\tiny BH}}}
\newcommand{\NS}{{\mbox{\tiny NS}}}

\newcommand{\MG}{{\mbox{\tiny MG}}}
\newcommand{\BD}{{\mbox{\tiny BD}}}

\begin{document}
\title {Gravitational Wave Tests of General Relativity with the
Parameterized Post-Einsteinian Framework}

\author{Neil Cornish}
\affiliation{Department of Physics, Montana State University, Bozeman, MT 59717, USA.}

\author{Laura Sampson}
\affiliation{Department of Physics, Montana State University, Bozeman, MT 59717, USA.}

\author{Nicol\'as Yunes}
\affiliation{Department of Physics, Montana State University, Bozeman, MT 59717, USA.}
\affiliation{Department of Physics and MIT Kavli Institute, 77 Massachusetts Avenue, Cambridge, MA 02139, USA.}

\author{Frans Pretorius}
 \affiliation{Department of Physics, Princeton University, Princeton, NJ 08544, USA.}

\date{\today}

\begin{abstract}
Gravitational wave astronomy has tremendous potential for studying extreme astrophysical phenomena and
exploring fundamental physics. The waves produced by binary black hole mergers will
provide a pristine environment in which to study strong field, dynamical gravity.
Extracting detailed information about these systems requires accurate theoretical models of the
gravitational wave signals. If gravity is not described by General Relativity,
analyses that are based on waveforms derived from Einstein's field equations could result in
parameter biases and a loss of detection efficiency.  A new class of ``parameterized
post-Einsteinian'' (ppE) waveforms
has been proposed to cover this eventuality. Here we apply the ppE approach to simulated
data from a network of advanced ground based interferometers (aLIGO/aVirgo) and from a future
space based interferometer (LISA). Bayesian inference and model selection are
used to investigate parameter biases, and to determine the level at which departures from general
relativity can be detected. We find that in some cases the parameter biases from assuming the wrong
theory can be severe. We also find that gravitational wave observations will beat the existing
bounds on deviations from general relativity derived from the orbital decay of binary pulsars
by a large margin across a wide swath of parameter space.
\end{abstract}

\maketitle

\section{Introduction}
Einstein's theory of gravity has been subject to a wide array of experimental tests and has
passed them all with flying colors~\cite{lrr-2006-3}. None of these tests, however, has probed
the strong field, dynamical regime that pertains to the final inspiral and merger of compact
objects. The Hulse-Taylor
binary pulsar PSR B1913+16~\cite{Hulse:1974eb} and the double binary pulsar PSR J0737-3039A~\cite{Burgay:2003jj,Kramer:2006nb} have provided convincing
evidence for the existence of gravitational waves, and have served as unique laboratories to test
general relativity (GR), but these objects have relatively small orbital velocities, $v/c \sim 10^{-3}$,
a mere factor of 10 faster than the Earth's orbit around the Sun. The parameter space covered by
black hole mergers, where orbital velocities $v/c \gg 10^{-3}$ and can approach $v/c \sim 0.7$, 
is currently {\em terra incognita} - Dragons may yet lurk there.

If not accounted for, the possibility that Einstein's theory of gravity may not correctly describe
the production and propagation of gravitational waves could have dire consequences for gravitational
wave astronomy. In the case of ground-based detectors, the detection of weak signals buried below
the instrument noise requires accurate models of the gravitational waveforms. Errors in the modeling
of these waveforms can lead to a loss in detection efficiency. When the signals are stronger, as will
often be the case with space-based observations of black hole mergers, waveform templates will no
longer be needed for detection, but a waveform model will be required to infer the physical parameters
of the system, such as the masses and spins of the black holes, and the
distance to the system. Waveform models based on an incorrect theory of gravity will lead to
{\em fundamental bias}~\cite{Yunes:2009ke} in the recovered parameters. Because these waveforms would not accurately describe nature, the parameters that maximize the fit of such a waveform to data would not correspond to the true physical values of the system. This bias is distinct from that caused by imperfect modeling of GR, as explored in~\cite{Cutler:2007mi}, as it reflects a fundamental lack of knowledge about the true nature of gravity, and not simply the use of inaccurate physical assumptions -- see~\cite{Yunes:2009ke} for more details.

Turning the problem around, the discovery that Einstein's theory is flawed would be the greatest result
to come out of gravitational wave astronomy~\cite{Schutz:2009tz}. This has served as the motivation for the development of
a wide range of tests of GR that use gravitational wave observations.
These tests can be broadly classified as ``extrinsic'' or ``intrinsic''. Extrinsic tests are possible
when there is a concrete alternative theory, such as massive gravitons~\cite{Will:1997bb,Will:2004xi,Berti:2005qd,Stavridis:2009mb,Arun:2009pq,Keppel:2010qu,Yagi:2009zm}, or Brans-Dicke
theory~\cite{Will:1994fb,Scharre:2001hn,Will:2004xi,Berti:2005qd,Yagi:2009zm}. Intrinsic tests work within the confines of GR, and take
the form of internal consistency checks, such as measuring the multipolar structure of the
metric~\cite{PhysRevD.52.5707,Collins:2004ex}, or multi-modal spectroscopy of BH inspiral and ringdown
waveforms~\cite{Berti:2005ys,Berti:2007zu}. These tests are valuable, but they do not cover the full
spectrum of possibilities. The existing extrinsic tests are limited by the lack of viable alternative
models, while the intrinsic tests do not so much test GR, as ``test the nature of massive compact
bodies within GR'' (to quote~\cite{Hughes:2006pm}).

Convincing alternative models to GR are hard to find because none of the currently proposed alternatives can satisfy 
key criteria that physicists would like to require. On the observational front, one wishes that any GR alternative 
passes all Solar System and binary pulsar tests with flying colors, only predicting deviations from GR in the strong-field 
regime, where tests are currently lacking. Many theories, such as Brans-Dicke 
theory~\cite{Will:1994fb,Scharre:2001hn,Will:2004xi,Berti:2005qd,Yagi:2009zm}, are heavily constrained 
by this requirement~\cite{lrr-2006-3}. On the theoretical front, one would wish viable GR alternatives to lead to well-posed
theories, with a positive definite Hamiltonian and free of instabilities. All perturbative string theory and loop quantum gravity 
low-energy effective theories~\cite{Alexander:2009tp,Yunes:2011we} currently lead to higher-derivative theories, which 
might violate this theoretical criteria. 

The paucity of concrete alternative models to GR~\cite{2010GWN.....4....3S} has impacted other testing grounds,
such as those based on solar system observations, or the aforementioned binary pulsar systems. In
those instances the standard approach has been to develop models that parameterize a wide class
of possible departures from GR - the parameterized post-Newtonian formalism~\cite{Nordtvedt:1968qs,1971ApJ...163..611W,1972ApJ...177..757W,1972ApJ...177..775N} and the
parameterized post-Keplerian formalism~\cite{1992PhRvD..45.1840D}. It is natural to adopt the same
strategy when analyzing
gravitational wave data, which leads to the parameterized post-Einsteinian (ppE) formalism
introduced in Ref.~\cite{Yunes:2009ke}. 

To motivate this approach, consider the standard post-Newtonian (PN) expression
for the dominant contribution to the stationary phase waveform describing the Fourier transform of the time-domain gravitational wave strain signal of the inspiral of two
non-spinning black holes on circular orbits (see e.g.~\cite{Berti:2005qd}): 
\begin{equation}\label{GR}
\tilde{h}_{\rm GR}(f) = \sqrt{\frac{5}{24}} \frac{{\cal{C}}}{\pi^{2/3}} {\cal{A}}(f) \frac{{\cal{M}}^{5/6}}{D_{L}} e^{i \Psi(f)} \, ,
\end{equation}
where $f$ is frequency,  ${\cal M} = \eta^{3/5} M$ is the chirp mass, $M=m_1+m_2$ is the total mass, $\eta = m_1 m_2/M^2$
is the dimensionless, symmetric mass ratio, $D_L$ is the luminosity distance and ${\cal{C}}$ is a geometric factor that depends on the relative orientation of the binary and the detector (its average for LISA is $\bar{\cal{C}} = 2/5$). The amplitude ${\cal{A}}(f)$ and
phase $\Psi(f)$ are developed as a series in $u = \pi {\cal M} f= \eta^{3/5} v^3$, where $v$ is the relative velocity between the two bodies~\cite{Yunes:2009yz} :
\begin{equation}\label{amp}
{\cal{A}}(f) = \sum_{k=0}^\infty \gamma_k u^{(2k-7)/6} \, .
\end{equation}
and
\begin{equation}\label{phase}
\Psi(f) = 2\pi f t_c - \Phi_c + \sum_{k=0}^\infty \left[ \psi_k +\psi_{kl}\ln u\right] u^{(k-5)/3} \, .
\end{equation}
The coefficients $\gamma_k(\eta)$, $\psi_k(\eta)$ and $\psi_{kl}(\eta)$ are currently known up
to $k=7$ in the post-Newtonian expansion of GR. 

In the simplest proposal of Yunes and Pretorius~\cite{Yunes:2009ke}, the phase and amplitude are modified by only one ppE term each, but as pointed out by the authors there is no reason to believe that an alternative theory of gravity will predict such a restricted deviation from GR. In view of this, Yunes and Pretorius proposed four different parameterizations that differed in their level of complexity, one of the most complicated of which is (see Eq. (46) in~\cite{Yunes:2009ke})
\begin{eqnarray}
&& {\cal{A}}(f) \rightarrow \left(1+\sum_i \alpha_i u^{a_i}\right) A_{\rm GR}(f) \, , \nonumber \\
&& \Psi(f) \rightarrow  \left(\Psi_{\rm GR}(f)+\sum_i \beta_i u^{b_{i}}\right) \, ,
\end{eqnarray}
where the coefficients $\alpha_i$ and $\beta_i$ may depend on the symmetric mass ratio $\eta$ (and in more general cases, also on the spin angular momenta and the difference between the two masses) and
$ A_{\rm GR}$ and $\Psi_{\rm GR}$ are the standard expressions in Eqs.~(\ref{amp}) and (\ref{phase}).
This is in essence the ppE approach.  

In an earlier study, Arun {\it et.al.}~\cite{Arun:2006hn,Arun:2006yw,Mishra:2010tp} considered
what can now be interpreted as a restricted version of the ppE formalism in which the
exponents $a_i$ and $b_i$ are required
to match those found in GR. This amounts to asking
how well the standard PN expansion coefficients could be recovered from gravitational wave
observations. They also developed internal self-consistency checks based on the observation
that each coefficient $\psi_k(\eta)$ provides an independent estimate of the mass ratio $\eta$.
While interesting, these tests are limited in scope as few of the well known alternative
theories of gravity (Brans-Dicke~\cite{Will:1994fb,Scharre:2001hn,Will:2004xi,Berti:2005qd,Yagi:2009zm}, Massive Graviton~\cite{Will:1997bb,Will:2004xi,Berti:2005qd,Stavridis:2009mb,Arun:2009pq,Keppel:2010qu,Yagi:2009zm}, Chern-Simons~\cite{Alexander:2007:gwp,Yunes:2010yf,Yunes:2009hc,Sopuerta:2009iy,Alexander:2009tp}, Variable $G$~\cite{Yunes:2009bv},  TeVeS~\cite{Bekenstein:2004ne}{\it etc.})
have corrections with exponents $a_i$ and $b_i$ that match those of GR~\cite{Yunes:2009ke}.
The full ppE formalism allows us to look for a much wider and realistic set of possible departures from GR.

Our goal here is to study how the ppE formalism can be used to search for waveform deviations from GR
 using data from the next generation of ground based interferometers (aLIGO/aVirgo) and
future space based interferometers ({\it e.g.} LISA). Bayesian model selection is used to
determine the level at which departures from GR can be detected (See Ref.\cite{2011arXiv1101.1391D}
for a related study that uses Bayesian inference to study constraints on Massive Graviton theories).
Advanced Markov Chain Monte Carlo (MCMC) techniques are
used to map out the posterior distributions for the models under consideration. From these
distributions, we are able to quantify the degree of fundamental bias in parameter extraction,
and in particular, if the fundamental bias can be significant in situations where there is
no clear indication that there are departures from GR. 

Recently, Pozzo et.al. [37] performed a similar study that applied Bayesian
model selection to estimate the bounds that could be placed on
massive graviton theory. As such, their work is a sub-case of the ppE
framework, {\it i.e.} a particular choice of $(b, \beta)$. Their implementation differed
from ours in that they used Nested Sampling while we used MCMC techniques,
but as we will show, our results are in agreement with theirs for the relevant
sub-case.

We find that gravitational wave observations will allow us to extend the existing
bounds derived from pulsar orbital decay~\cite{2010PhRvD82h2002Y} into the region of parameter
space that covers strong field departures from GR ($a_i>0$ and $b_i > -5/3)$ 
(see Fig.~\ref{beta-bounds}--\ref{alpha-bounds} in Sec.~\ref{comp-boundes}). 
As expected, we find that
the strength of the bounds on the ppE parameters are
inversely proportional to the signal-to-noise ratio (SNR),
and the extent to which deviations between GR templates
and non-GR signals can be detected (the departure of the
``fitting factor'' from unity) scales as $1/{\rm SNR}^2$.
The logarithm of the odds ratio used
to decide if a signal is described by GR or some alternative theory follows the
same $1/{\rm SNR}^2$ scaling. A more surprising result is the possibility of ``stealth bias''
whereby the parameters recovered using GR templates can be significantly biased even when
the odds ratio shows no clear preference for adopting an alternative theory of gravity. 

The remainder of this paper is organized as follows.  Section~\ref{sec:analysis-framework} introduces
the analysis framework in more detail, including a discussion
of the waveform model, noise spectrum, and Bayesian tools used.  
Section~\ref{sec:computational-techniques} describes in detail the computational techniques
used to to implement the analysis.
Section~\ref{sec:results} presents the results of our analysis.
Section~\ref{sec:Conclusions} closes with a discussion of how our results might change as the
degree of realism is increased, and identifies key questions to be addressed in future work.
Throughout this paper we use geometric units with $G=c=1$. 

\section{Analysis Framework}
\label{sec:analysis-framework}

\subsection{Bayesian Inference}
Questions of model selection and parameter biases can be addressed very naturally in the framework of
Bayesian inference. This approach is now well established in the field of gravitational wave data
analysis, as are the tools used to carry out the analysis. To avoid unnecessary repetition, we will
focus on those aspects of the analysis that are new, and refer the reader to
Ref.~\cite{Littenberg:2009bm} for a detailed description of the techniques used.

We are interested in comparing the hypothesis $\mathcal{H}_0$ that gravity is described by GR with the
hypothesis $\mathcal{H}_1$ that gravity is described by an alternative
theory belonging to the ppE class. Here we are dealing with nested hypotheses, as the ppE models
include GR as a limiting case. When new data $d$ becomes available, our prior belief
$p(\mathcal{H})$ in hypothesis $\mathcal{H}$ is updated to give the posterior belief $p(\mathcal{H}|d)$.
Bayes' theorem tells us that
\begin{equation} \label{Bayes' theorem}
p(\mathcal{H}|d)=\frac{p(d|\mathcal{H})p(\mathcal{H})}{p(d)}\, ,
\end{equation}
where $p(d|\mathcal{H})$ is the (marginal) likelihood of observing the data $d$ if the hypothesis holds, and $p(d)$
is a normalization constant. For hypotheses described by models with continuous parameters,
the likelihood $p(d|\mathcal{H})$ is found by marginalizing the likelihood $p(d|\vec{\theta},\mathcal{H})$
of observing data $d$ for model parameters $\vec{\theta}$:
\begin{equation} \label{evidence_integral}
p(d|\mathcal{H}) = \int d\vec{\theta} \ p(\vec{\theta},\mathcal{H})p(d|\vec{\theta},\mathcal{H}) \, ,
\end{equation}
where $p(\vec{\theta},\mathcal{H})$ is the prior distribution of the parameters. The marginalized
likelihood, $p(d|\mathcal{H})$, is also known as the evidence for a given model. Hypotheses are
compared by computing the odds ratio, or Bayes factor:
\begin{equation}
BF = \mathcal{O}_{1,0} \equiv \frac{p(\mathcal{H}_1|d)}{p(\mathcal{H}_0|d)}
=\frac{p(\mathcal{H}_1)}{p(\mathcal{H}_0)}\frac{p(d|\mathcal{H}_1)}{p(d|\mathcal{H}_0)} \, ,
\label{BF-def}
\end{equation}
which gives the ``betting odds'' of $\mathcal{H}_1$ being a better description of Nature than
$\mathcal{H}_0$. The normalization constant $p(d)$ cancels in the odds-ratio. The prior odds
ratio $p(\mathcal{H}_1)/p(\mathcal{H}_0)$ gets updated by the likelihood ratio, $p(d|\mathcal{H}_1)/p(d|\mathcal{H}_0)$, which is also known as the evidence ratio. In Bayesian analysis
``today's posterior is tomorrow's prior''~\cite{lindley}, and $p(\mathcal{H}|d)$ is used in
place of $p(\mathcal{H})$ in subsequent analyses.
While a single black hole inspiral event may not yield strong evidence for a departure from
GR, several such observations can be combined to make a more compelling case.

In addition to simply detecting deviations from GR, we are also interested in studying how departures from GR might affect parameter estimation.
This can be assessed by looking at the posterior distribution function $p(\vec{\theta} |d, \mathcal{H})$,
which describes the probability distribution for parameters $\vec{\theta}$ under the
assumption that the signals are described by model $\mathcal{H}$ given data $d$. The posterior
distribution is given by the product of the prior and the likelihood, normalized by the evidence:
\begin{equation} \label{Bayes2}
p(\vec{\theta}|d, \mathcal{H})=\frac{p(\vec{\theta}, \mathcal{H}) p(d|\vec{\theta}, \mathcal{H})}{p(d | \mathcal{H})} \, .
\end{equation}
Once the prior distribution and the likelihood function have been specified we are left with the purely
mechanical task of computing the posterior distributions and odds ratio for competing hypotheses.

\subsection{Waveform Model}

The original ppE waveforms were for non-spinning, equal mass binaries in quasi-circular orbits, and
included a description of the dominant harmonic through inspiral, merger and ringdown.  In the current
analysis we restrict our attention to the inspiral portion of the waveform, but our signals come from unequal mass binaries.  We have examined the generalization of the ppE framework for unequal mass systems, and find that for a single detection it is indistinguishable from the equal mass case. Including multiple detectors, and the merger and ringdown phases, which increase the signal-to-noise ratio, can help break parameter degeneracies that exist in the inspiral phase, but these benefits come at the cost of having to consider additional ppE parameters. We will consider this in a separate publication.

In the stationary phase approximation, our ppE waveforms are parameterized as follows
\begin{equation} \label{ppEwaveform1}
\tilde{h}(f) = \tilde{h}_{\rm GR}(f)\left[ 1 + \alpha \; u^a\right] e^{i \beta \, u^b} \quad f < f_{\rm max} \,,
\end{equation}
where $(\alpha,a)$ are amplitude ppE parameters and $(\beta,b)$ are phase ppE parameters. As noted previously, both $\alpha$ and $\beta$ can depend on the spin angular momenta and mass difference of the two bodies, as well as the symmetric mass ratio of the system. With a single detection, however, these dependencies are impossible to determine, and so we defer an analysis of them to future work.
Here $\tilde{h}_{\rm GR}(f)$ is the usual GR waveform quoted in Eq.~(\ref{GR}). We set the maximum frequency cut-off 
at twice the innermost stable circular orbit frequency of a system described by GR. A more
consistent choice would be to use the minimum of the ppE energy function, but the results were
found to be fairly insensitive to the choice of $f_{\rm max}$. To simplify the analysis we restrict
our attention to the lowest PN order in the amplitude of Eq.~(\ref{amp}), setting $\gamma_k=0$ for $k>0$.
The GR phase terms in Eq.~(\ref{phase}) are kept out to $k=7$.
Furthermore, we limit the range of the ppE parameters $a$ and $b$ to not be greater
than these corresponding highest order PN terms, namely $a < 2/3$ and
$b < 1$.
\footnote{It is certainly conceivable
that the {\em leading order} deviation arising from an alternative theory comes in at some
high order, and has a much larger magnitude than the nearest exponent term
in the PN expansion. Thus it is not {\em a priori} inconsistent to allow a range of exponents outside
of that of the PN expansion used for the GR signal in the ppE waveforms, though this would
require more complicated priors on the amplitudes, and so for simplicity in this study
we restrict to the stated range.}

As discussed in the Introduction, the ppE framework introduces $i$ sets of ppE theory parameters
$(\alpha_{i}, a_{i}, \beta_{i}, b_{i})$ that modify the amplitude and phase, but we here work to 
leading order, keeping only the $i=0$ set. This approach will tend to over-estimate how well the 
ppE parameters $(\alpha_{0}, a_{0}, \beta_{0}, b_{0}) \equiv (\alpha_{}, a_{}, \beta_{}, b_{})$
can be constrained by the data. A better approach, which we intend to pursue in future studies, is to marginalize
over the higher order terms. 

Table~I lists the leading ppE corrections that have been computed
for several alternative theories of gravity. Generally, the exponents $a$ and $b$ are pure numbers fixed by the theory,
while the amplitudes $\alpha$ and $\beta$ are free parameters that relate to the unknown coupling strengths
of the modified/additional gravitational degrees of freedom.

\begin{table}[tp] 
\begin{ruledtabular}
\begin{tabular}{c|c|c|c|c}
Theory     & $\; a \;$	& $\; \alpha\;$	& $\; b\;$   &  $\; \beta\;$	\\ \hline
Brans-Dicke~\cite{Will:1994fb,Scharre:2001hn,Will:2004xi,Berti:2005qd,Yagi:2009zm}& --	        & 0	        & -7/3	& $\beta$	\\
Parity-Violation~\cite{Alexander:2007:gwp,Yunes:2010yf,Yunes:2009hc,Sopuerta:2009iy,Alexander:2009tp}& 1	        & $\alpha$      & 0	& --      	\\
Variable $G(t)$~\cite{Yunes:2009bv} & 	-8/3    & $\alpha$	& -13/3	& $\beta$	\\
Massive Graviton~\cite{Will:1997bb,Will:2004xi,Berti:2005qd,Stavridis:2009mb,Arun:2009pq,Keppel:2010qu,Yagi:2009zm} & 	--    & 0	& -1	& $\beta$	\\
Quadratic Curvature~\cite{Stein:2010pn,Yunes:2011we} & 	--    & 0	& -1/3	& $\beta$	\\
Extra Dimensions~\cite{Yagi:2011yu} & 	--    & 0	& -13/3	& $\beta$	\\
Dynamical Chern-Simons~\cite{DCS} & +3    & $\alpha$    & +4/3  & $\beta$\\
\end{tabular}
\end{ruledtabular}
\caption{\label{tab1} Leading ppE corrections in several alternative theories of gravity
(GR corresponds to $\alpha=\beta=0$). In dynamical Chern-Simons gravity, $(\alpha,\beta)$ are proportional to the spin-orbital angular momentum coupling. For non-spinning binaries,
the last row would simplify to $(\alpha,\beta) = (0,0)$, but we include it here for completeness.}
\end{table}

\subsection{Instrument Response}

The aLIGO/aVirgo analysis was performed using simulated data from the 4 km Hanford and Livingston detectors
and the 3 km Virgo detector. The time delays between the sites and the antenna beam patterns were computed
using the expression quoted in Ref.~\cite{Anderson:2002lg}. Since the detectors barely move relative to the
source during the time the signal is in-band, the antenna patterns can be treated as fixed and the
time delays $\Delta t$ between the sites can be inserted as phase shifts of the form $2\pi f \Delta t$.
For the instrument noise spectral density, we assumed all three instruments were operating in a wide-band
configuration with
\begin{equation}
S_n(f) = 10^{-49}\left( x^{-4.14}-5 x^{-2}+111 \frac{(2-2 x^2+x^4)}{2+x^2} \right) \, ,
\end{equation}
and $x = (f/215 {\rm Hz})$.

The space based (LISA) analysis was performed using the $A$ and $E$ Time Delay Interferometry
channels~\cite{Prince:2002hp} in the low frequency approximation~\cite{Cornish:2002rt,Rubbo:2003ap}. It is known
that this approximation can lead to biases in some of the recovered parameters, such as polarization and inclination
angles. This, however, is an example of a modeling bias introduced by inaccurate physical assumptions, and not of a
fundamental bias resulting from incomplete knowledge of the theory describing gravity. In our current study the
modeling bias is avoided by using the same low frequency response model to produce the simulated data and to
perform the analysis.

In contrast to the ground
based detectors, the signals seen by LISA are in-band for an extended period of time, and the motion
of the detector needs to be taken into account. The time dependent phase delay between the detector
and the barycenter and the time dependent antenna pattern functions are put into a form that can
be used with the stationary phase approximation waveforms by mapping between time and frequency
using $t(f) = (d \Phi/ d f)/2\pi$. Details of this procedure can be found in Ref.~\cite{Cutler:1997ta}.
The noise spectral density model includes instrument noise and an estimate of the foreground confusion
noise from unresolved galactic binaries, matching those quoted in Ref.~\cite{Key:2010tc}.

\subsection{Likelihood Function}

Under the assumption that the noise is Gaussian, the likelihood that the data $d$ would arise from a
signal with parameters $\vec{\theta}$ is given by
\begin{equation}
p(d\vert \vec{\theta}) = C e^{-\chi^2(\vec{\theta})/2} \, ,
\end{equation}
where $C$ is a constant that depends on the noise level. Here 
\begin{equation}\label{chisq}
\chi^2(\vec{\theta}) = (d-h(\vec{\theta})\vert d-h(\vec{\theta})) \, ,
\end{equation}
and the brackets denote the noise weighted inner product
\begin{equation}\label{nwip}
(a\vert b) = 2 \int \frac{\tilde{a}(f) \tilde{b}^*(f) + \tilde{a}^*(f) \tilde{b}(f)}{S_n(f)} \, df \, .
\end{equation}
For a theoretical study that assumes the noise is Gaussian and has a known spectrum, there is no need to
add simulated noise to the data - the appropriate spread in the parameter values and overall topography of
the likelihood surface follow from the functional form of the signal and the noise weighting in Eq.~(\ref{nwip}). Thus, we may write $d = h(\vec{\theta}')$ where $\vec{\theta}'$ are the true source
parameters.

Many alternative theories of gravity predict the existence of polarization states beyond the
usual ``plus'' and ``cross'' polarizations of GR that complicate the treatment of the instrument
response, whose Fourier transform is
\begin{eqnarray}
\tilde{h}_{inst} &=&  F_{+} \tilde{h}_{+} + F_{\times} \tilde{h}_{\times} + F_{S} \tilde{h}_{S} \nonumber \\
 && + F_{L} \tilde{h}_{L} + F_{V1}  \tilde{h}_{V1} + F_{V2} \tilde{h}_{V2} \,,
\end{eqnarray}
Here $\tilde{h}_{+\times}$ are the usual plus and cross-polarization states, $\tilde{h}_{S}$ is a scalar (breathing) mode, $\tilde{h}_{L}$ is a scalar longitudinal model and $\tilde{h}_{V1,V2}$ are two vectorial modes~\cite{Cliff-Eric-Book}, while the $F$'s are the  detector antenna patterns~\cite{Will:1993ns}, which depend on the
sky location $(\theta,\phi)$ and polarization angle $\psi$ of the signal. 

To simplify the analysis we assume the usual polarization content for a circular binary
viewed at inclination angle $\iota$ and neglect the other contributions:
\begin{eqnarray}
\tilde{h}_+ &=& (1+\cos^2\iota) \Re(\tilde{h})+ 2\cos\iota\, \Im(\tilde{h}) \, , \nonumber \\
\tilde{h}_\times &=& (1+\cos^2\iota) \Im(\tilde{h}) - 2\cos\iota\, \Re(\tilde{h}) \, .
\end{eqnarray}
In other words, we have assumed that the signal
in the detector has the form $\tilde{s}(f) = F(\theta,\phi,\psi,\iota) \; \tilde{h}(f)$ with the function
$F(\theta,\phi,\psi,\iota)$ given by the usual GR expression. If additional polarization
states were present, this assumption would result in a reduction in detection
efficiency and biases in the recovery of the extrinsic parameters $(\theta,\phi,\psi,\iota)$.

The justification for making this simplification is that we are primarily interested in
how well the intrinsic parameters $(\alpha, a, \beta, b)$ can be constrained, and
we expect these parameters to be only weakly correlated with the extrinsic parameters.
The presence of additional polarization states will provide an additional handle on detecting
departures to GR~\cite{1978PhRvD17.3158H, 2008ApJ685.1304L, Tinto:2010hz}, and we plan to
explore this possibility in the context of the ppE formalism in future work.

Defining $A_+ = |F^+ \tilde{h}_+(f;\vec{\theta})|$ and $A_\times = |F^\times \tilde{h}_\times(f;\vec{\theta})|$, and similarly for
$\vec{\theta}'$, the chi-squared goodness of fit of Eq.~(\ref{chisq}) can be re-expressed as
\begin{eqnarray}
\chi^2(\vec{\theta}) &=&  4 \int \frac{df}{S_n(f)}\left[A_+^2+A_\times^2+{A'}_+^2+{A'}_\times^2
\right.
\nonumber \\
&-& \left.2 (A_+ {A'}_+ + A_\times {A'}_\times)\cos\Delta \Psi  
\right.
\nonumber \\
&-& \left. 
2 (A_\times {A'}_+ - A_+ {A'}_\times)\sin\Delta \Psi \right],
\end{eqnarray}
where $\Delta \Psi = \Psi(\vec{\theta}) - \Psi(\vec{\theta}')$. As noted in Ref.~\cite{Cornish:2010kf}, in
the regime of interested where $\chi^2$ is small, all the terms in the above integrand are slowly
varying functions of frequency, so it is possible to compute the likelihood very cheaply using an
adaptive integrator.

\subsection{Priors}

As we shall see, the choice of priors on the ppE parameters has a significant effect on the results,
especially when it comes to model selection. The natural priors on the ppE parameters are those that
come from existing data on binary pulsars, but these turn out to range from very restrictive to wide
open depending on what sector of the ppE parameter space is being examined. To simplify the analysis we adopt
uniform priors for the ppE parameters and
seek to determine where direct GW observations would prove more constraining than the existing
binary pulsar observations. 

The priors on the exponents $a$ and $b$ are taken to be uniform across the
ranges $a\in [-3,2/3]$ and $b\in [-4.5,1]$. The upper end of the range is chosen so that the ppE corrections
to the amplitude and the phase do not go to higher order in the expansion parameter $u$ than the
post-Newtonian order of the reference GR waveforms.
The lower end of the range is chosen to cover
all known alternative theories, though in any case, the low end of the range turns out to be far better
constrained by binary pulsar observations. 

The priors on $\alpha$, and $\beta$ are more difficult to set. Lacking any theoretical or experimental guidance, we assign
uniform priors for the amplitudes $\alpha,\beta \in [-1000,1000]$. The range in $\alpha, \beta$ is set such that it is sufficiently large that at the most positive end of the prior ranges on $a, b$, the exploration of possible values of $\alpha, \beta$ is not restricted by prior bounds. That is, even in the most poorly-constrained region of the ppE parameter-space, the constraints are not due to an overly restrictive prior. 

The parameters used to describe the black hole binary were the log of the total mass $M$ and the log of
the chirp mass ${\cal M}$, the sky location $(\cos \theta, \phi)$, orbital plane orientation
$(\cos \theta_L,\phi_L)$, merger phase $\Phi_c$, merger time $t_c$, and luminosity distance $D_L$.
The angular parameters are taken to have uniform priors that covered their natural range. For the
aLIGO studies, we assign uniform priors: $\ln(M/M_\odot) \in [1.3,5.3]$;
$\ln({\cal M}/M_\odot) \in [0.55,4.5]$; $t_c/{\rm s} \in [1,16]$; $D_L/{\rm Mpc} \in [0.1,10^4]$.
For the LISA studies, we assign uniform priors: $\ln(M/M_\odot) \in [12.2,16.8]$;
$\ln({\cal M}/M_\odot) \in [11.4,16]$; $t_c/{\rm s} \in [1,6\times 10^7]$; $D_L/{\rm Gpc} \in [0.01,1000]$.
While we could use more physically motivated priors for the black hole parameters (such as distance
priors that scaled with $D_L^2$), these choices have little effect on the model comparison between GR and
ppE waveforms.

\section{Computational Techniques}
\label{sec:computational-techniques}

Posterior distribution functions for the alternative hypotheses were computed using the Markov Chain
Monte Carlo (MCMC) implementation described in Ref.~\cite{Littenberg:2009bm}, additionally enhanced
by adding Differential Evolution~\cite{TerBraak:2006:MCM:1145406.1145416,TerBraak:2008:DEM:1484982.1484994}
to the mix of proposal distributions. The evidence for the competing hypotheses was calculated using
the volume tessellation algorithm~\cite{Weinberg:2009rd} and cross-checked using thermodynamic
integration~\cite{goggans:59}.

The ppE waveforms introduce a number of complications that make parameter estimation and model
selection challenging. These complications can be seen when using the quadratic Fisher matrix
approximation $\Gamma_{ij} = -\partial_i\partial_j \langle \ln p(\vec{\theta}\vert d) \rangle$
to estimate the parameter correlation
matrix $C^{ij} = \langle \Delta \theta^i \Delta \theta^j \rangle\approx \Gamma_{ij}^{-1}$.
When evaluated at the GR limit point $(\alpha,\beta)=(0,0)$, the quadratic approximation to the
Fisher matrix is singular, and it is necessary to include higher order derivatives to
obtain a finite covariance matrix. The situation is worse when $a=0$, as then
$\alpha$ is fully degenerate with $D_L$, and when $b=0$, as then $\beta$ is fully degenerate
with $\Phi_c$. Partial degeneracies also exist whenever the $a$ or $b$ exponents match the
exponents found in the post-Newtonian expansion of GR. 

The various degeneracies and parameter correlations do not constitute a fundamental problem with
the ppE formalism, but they do demand that we use very effective MCMC samplers that are able to
fully explore the parameter space. The algorithm described in Ref.~\cite{Littenberg:2009bm} uses
parallel tempering with multiple, coupled chains, with each chain exploring a tempered likelihood
surface $p(d\vert \vec{\theta})^{1/T}$. The high temperature chains explore more widely, and can
communicate this information via parameter exchange to the $T=1$ chain that is used for
parameter estimation. Parallel tempering helps the Markov chains explore complicated posterior
distributions, but convergence can still be slow if the proposal distributions are not well chosen.

The ultimate proposal distribution is the posterior distribution itself, but since that is unavailable
in advance, we have to make do with approximations to this ideal. The covariance matrix $C_{ij}$
provides a local approximation to the posterior distribution. It can be estimated semi-analytically
using the Fisher information matrix, or more directly from the recent past history of the Markov chain
itself. The latter approach introduces hysteresis into the chains, but so long as the covariance matrix
is only updated occasionally the chains are asymptotically Markovian.  
In the present study, we continued to use the Fisher matrix based proposal
distributions described in Ref.~\cite{Littenberg:2009bm}, but found that the convergence time of the chains was very long until we augmented these techniques with proposals based on Differential Evolution.

Differential Evolution (DE) provides an approximation to the posterior distribution based on the
past history of the chains. Unlike methods based on the covariance matrix, DE works extremely well
with highly correlated parameters. In its original formulation,
DE~\cite{TerBraak:2006:MCM:1145406.1145416} was designed
to work with a population of $N$ parallel chains (all with temperature $T=1$).
The idea is very simple and can be coded in a few lines: Chain $i$ is updated
by randomly selecting chains $j$ and $k$ with $j \neq k \neq i$, forming
the difference vector $\vec{\theta}_j-\vec{\theta}_k$ and proposing the move
\begin{equation}
\vec{y}_i = \vec{\theta}_i + \gamma(\vec{\theta}_j-\vec{\theta}) \, .
\end{equation}
For $D$-dimensional multivariate normal distributions, the optimal choice
for the scaling is $\gamma = 2.38/\sqrt{2 D}$. Since the difference vector points along the
$D$-dimensional error ellipse, the jumps are usually ``in the right direction.''
It is a good idea to occasionally (e.g.~10\% of the time) propose
jumps with $\gamma = 1$, which act as mode-hopping jumps when the samples $(j,k)$ come
from separate modes of the posterior.

The original formulation of DE is not very practical since it requires $N> 2D$ parallel chains
for each rung on the temperature ladder. A more economical approach is to use samples from the past
history of each chain~\cite{TerBraak:2008:DEM:1484982.1484994}. It can be shown that
this approach is asymptotically Markovian in the limit as one uses the full past history
of the chain. We have implemented a variant of the DE algorithm as follows:\\

$\bullet$ Create a history array for each parallel chain. Initialize a counter $M$.
Store every $~10^{\rm th}$ sample in the history array and add to the counter
each time a sample is added. DE moves are more effective if points during the burn-in phase of the search are discarded from the history array.\\

$\bullet$ Draw two samples from the history array: $j \in [1,M]$,
$k \in [1,M]$ and repeat if $k=j$. Propose the move to
\begin{equation}
\vec{y} = \vec{\theta} + \gamma(\vec{\theta}_j-\vec{\theta}_k) \, .
\end{equation}
Here we draw $\gamma$ from a Gaussian of width $2.38/\sqrt{2 D}$ for 90\%
of the DE updates and set $\gamma =1$ for the rest.\\

The standard DE proposal seeks to update all the parameters at once,
but it is often more effective to update smaller sub-blocks of
highly correlated parameters. We did this in $\sim$ 30\% of the DE proposals.

The fraction of all proposed moves that use DE is a tunable parameter. We
used 60\% DE proposals, 30\% Fisher matrix based proposals, 5\% draws from the
prior distribution and 5\% uniform draws with width $\sim 10^{-6}$ of the prior range.
Notice that even though the Fisher matrix might be singular in certain regions of the
parameter manifold, one can still propose jumps with it. In those regions, the proposed
jumps will not lead to a better likelihood, and will simply be rejected.

With the mix of proposal distributions described above, and using $\sim 10$ parallel chains
geometrically spaced with $T_{i+1} = 1.3 T_{i}$, our MCMC implementation converges quickly
to a stationary distribution. The chains are typically run for 500,000 samples, with the
first 100,000 discarded based on a conservative estimate of the burn-in length.

The marginal likelihood, or evidence, $p(d | \mathcal{H})$ is computed using independent
codes supplied by Martin Weinberg and Will Farr that implement Weinberg's volume tessellation
algorithm (VTA)~\cite{Weinberg:2009rd}. The VTA uses the posterior samples from the Markov chain
to assign probability to a partition of the sample space and performs the marginal likelihood integral
directly. The samples are partitioned using a kd-tree, and volume elements containing $m$ samples (we use
$m=32$ or $m=64$) are used to provide a discrete approximation to the
integral in Eq.~(\ref{evidence_integral}). The integrand in each volume element is approximated using either
the average posterior density (Farr's code) or the median posterior density (Weinberg's code) of
the $m$ samples in the volume element. The VTA is applied to a sub-sample of the full chain, and by
repeating the calculation with different subsamples in a process called bootstrapping, it is possible
to compute statistical errors bars on the evidence caused by using finite length Markov chains.

There is a trade-off in the choice of the boxing number $m$, with large values of $m$ providing
better estimates of the average or mean posterior density in each cell, and small values of $m$
providing better resolution to features in the posterior. In our experience, the statistical error
found from the bootstrap procedure is usually smaller than the systematic error that we estimate
by varying the boxing size from $m=16$ to $m=64$. 

As a cross check we applied thermodynamic integration~\cite{goggans:59} to a few test cases using the 
implementation described in the appendix of Ref.~\cite{Littenberg:2010gf}. In tests on distributions where 
the evidence can be calculated analytically, such as multi-variate Gaussians, we found that thermodynamic 
integration gave more accurate results. On the other hand, thermodynamic integration requires many more chains (upwards
of 50 for the ppE studies) and a careful tuning of the temperature ladder in order to resolve the
integrand. This tuning necessitates a long pilot run, or complicated adaptive tuning of the temperature
ladder. So while thermodynamic integration produces more accurate results, it requires careful
tuning and is far more computationally intensive. Based on the tests described in Appendix A, we estimate that the errors in the (natural) log Bayes factors computed using the VTA algorithm are of order $\pm 2$. 

\section{Results}
\label{sec:results}

We explore a range of questions concerning the application of the ppE formalism to
detecting departures from GR using gravitational wave observations from both LISA and
the three-detector network of aLIGO/aVIRGO interferometers. 
First, we derive simple estimates of how well the ppE
parameters can be constrained by gravitational wave data by using ppE templates to
detect GR signal injections. The spread in the recovered ppE parameters establishes
the range that is consistent with GR, and values outside of this range would
point towards a departure from GR. We then compare these simple
bounds to the more rigorous (and computationally expensive) bounds that can be
derived from Bayesian model selection. Finally, we explore how searching
for gravitational waves using GR templates can lead to biases in the recovered parameters
if Nature is described by an alternative theory of gravity. We find that these biases can
become significant before the evidence disfavors GR.

\subsection{Cheap Bounds and Comparison with Pulsar Bounds}
\label{comp-boundes}

The first question we seek to address in this paper is how well the four ppE parameters $(\alpha,a,\beta,b)$ can be determined.
One approach to answering this question is to examine how a search using ppE 
templates would look when used to characterize a signal that is consistent with
GR. That is, if the signal observed is described by GR to the given level of accuracy of our detectors, 
what values for the ppE parameters will be recovered from a search with ppE templates? 
Because we know that in GR the values of $\alpha$ and $\beta$ should be $0$ for all values of $a$ and $b$, we wish to determine
the typical spread in the recovered value of $(\alpha,\beta)$, centered at zero. The standard deviation in this spread
then gives us a constraint on the magnitude of the deviation that is still consistent with observations, ie.~deviations that
are `inside our observational error bars.' 

\begin{figure}[h]
\includegraphics[width=88mm,clip=true]{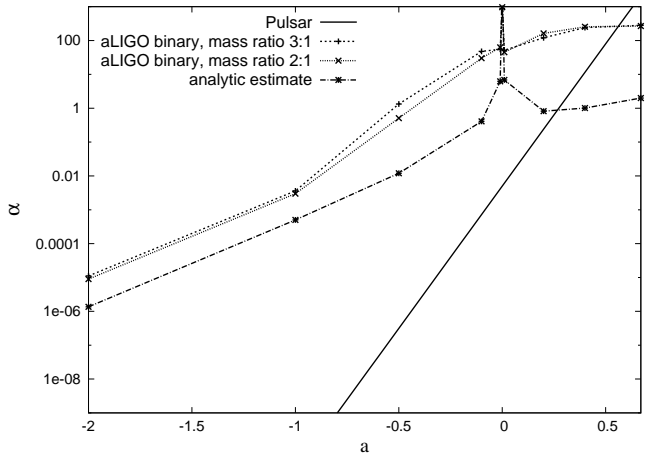}
\caption{\label{alpha-bounds} UPPER PANEL:Bounds on $\alpha$ for different values of $a$, found using two
different aLIGO sources. The two sources had different mass ratios, total masses, and sky locations, but were
scaled to have a network SNR of 20. The rough estimate for the $\alpha$ bound from
equation (\ref{rough}) is shown for comparison. Also included is the bound on $\alpha$ derived from the
golden pulsar (PSR J0737-3039) data. \newline
LOWER PANEL: Bounds on $\alpha$ for different values of $a$, found using two LISA sources at
redshift $z=1$ and $z=3$. The pulsar bound is shown for comparison. The sources injected had the same parameters as those from the lower panel in Figure ~\ref{beta-bounds} .}
\end{figure}

Cheap constraints will be defined as the $(3\sigma)$-bound on the posterior distribution of ppE parameters $\alpha$ or $\beta$,
while keeping $a$ or $b$ fixed and marginalizing over all other system parameters. 
These bounds are `cheap' because we do not
have to re-run a search with pure GR templates and then compute the evidence, via integration of the posterior, 
to compute the Bayes factor (the latter is particularly computationally expensive). These cheap bounds are similar to constraints studied
by looking at the $(\alpha,\alpha)$ or $(\beta,\beta)$ elements of the variance-covariance matrix. 
Our cheap constraints, however, are $3\sigma$ ones, in contrast to the more standard $1\sigma$ bounds quoted
from variance-covariance matrix studies.

Rough analytic estimates for the bounds on $(\alpha,\beta)$ can be derived by considering how the the ppE terms
affect the overall amplitude ${\cal A}$ and phase $\Psi$ of the signal:
\begin{eqnarray}
\Delta \ln{\cal A} &\simeq & \alpha (u_{\rm min}^a - u_{\rm max}^a) \nonumber \\
\Delta \Psi &\simeq & \beta (u_{\rm min}^b - u_{\rm max}^b).
\end{eqnarray}
Here $u_{\rm min}$ and $u_{\rm max}$ are the minimum and maximum values of the $u$ parameter. For the aLIGO sources
$u_{\rm min} \sim 3\times 10^{-3}$, while for the LISA sources $u_{\rm min} \sim 10^{-3}$. The ISCO cut-off in the
frequency evolution sets $u_{\rm max} \sim 3 \times 10^{-2}$ for moderate mass ratios. Combining these estimates with
a crude Fisher matrix estimate for how well the amplitude and phase are constrained: $\Delta \ln{\cal A} \sim 
\Delta \Psi \sim 1/{\rm SNR}$ yields the $3\sigma$ bounds
\begin{eqnarray}\label{rough}
\vert \alpha \vert  &\leq & \frac{3}{{\rm SNR}\, \vert u_{\rm min}^a - u_{\rm max}^a \vert} \nonumber \\
\vert \beta \vert &\leq& \frac{3}{{\rm SNR}\, \vert u_{\rm min}^b - u_{\rm max}^b \vert}.
\end{eqnarray}
These estimates reproduce the overall shape of the exclusion plots in the $(a,\alpha)$ and $(b,\beta)$ planes,
but they tend to over estimate the strength of the bounds as they do not take into account covariances with
other parameters. The $\alpha$ bounds turn out to be a factor of $\sim 10$ weaker due to covariances
between $\alpha$ and the distance and inclination, while the bounds on $\beta$ come out a factor of $\sim 100$
weaker due to covariances between $\beta$ and the chirp mass and mass ratio.

\begin{figure}[h]
\includegraphics[width=88mm,clip=true]{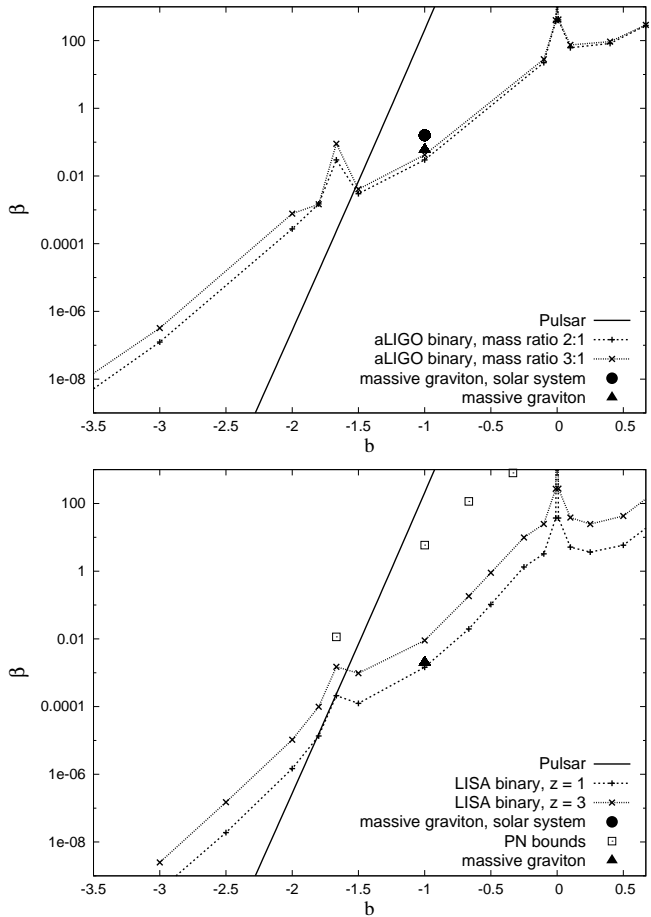}
\caption{\label{beta-bounds} UPPER PANEL: Bounds on $\beta$ for different values of $b$ for a
single ${\rm SNR}=20$ aLIGO/aVirgo detection. Plotted here is a $(3 \sigma)$ constraint, where $\sigma$
is the standard deviation of the $\beta$ parameter derived from the Markov chains. The sources injected had the
same parameters as those from the upper panel in  Figure ~\ref{alpha-bounds}.
Also included is the bound on $\beta$ derived from the golden pulsar (PSR J0737-3039) data, as well as bounds
found from solar system experiments and other aLIGO analyses for massive graviton theory. \newline
LOWER PANEL: Bounds on $\beta$ for different values of $b$ found using two LISA sources at
redshift $z=1$ and $z=3$. The pulsar bound is shown for comparison, as well as bounds found from solar system experiments and other LISA analyses for massive graviton theory. These other bounds are scaled to a system with $z=1$.}
\end{figure}

Figures~\ref{alpha-bounds} and \ref{beta-bounds} show these cheap constraints 
on the ppE amplitude parameters as a function of the exponents $a$ and $b$ for a variety of
aLIGO/aVirgo and LISA detections. 
To generate these plots, we injected GR signals and then searched on them with ppE templates.
For each search, either $a$ or $b$ was held fixed at a specific value, while the other three ppE
parameters (and all other system parameters) were allowed to vary. 
We then calculated the standard deviation of the posterior distribution of the relevant amplitude
parameter $\alpha$ or $\beta$, and used three times this value as the cheap bound shown on the
plots.  

A natural course of action might seem to be the following: marginalize over $a$ and $b$ as well, instead of keeping them fixed, and calculate constraints on $\alpha$ and $\beta$ this way. Looking at Figures~\ref{beta-bounds} and \ref{alpha-bounds}, however, show why this analysis would not be particularly helpful. The uncertainty in $\alpha$ and $\beta$ is so much higher at the positive ends of the prior ranges on $a$ and $b$ than at the negative ends that the Markov chains would spend almost all of their iterations exploring this area of parameter space if $a$ and $b$ were allowed to change. Thus, to get any knowledge about the uncertainties in $\alpha$ and $\beta$ for negative values of $a$ and $b$, we need to fix $a$ and $b$.

The aLIGO systems were chosen to have network ${\rm SNR}=20$,
but different masses and sky locations. One system had masses $m_1= 6 M_{\odot}$, $m_2= 18 M_{\odot}$ 
($\eta=0.1875$), $D_{L} = 258 \; {\rm{Mpc}}$, while the other had $m_1= 6 M_{\odot}$,
$m_2= 12 M_{\odot}$ ($\eta=0.2222$), $D_{L} = 462 \; {\rm{Mpc}}$. The LISA sources
were at different redshifts and had different masses and SNRs. The system at redshift $z=1$ had
$m_1= 1 \times 10^6 M_{\odot}$, $m_2= 3 \times 10^6 M_{\odot}$ ($\eta=0.1875$) and ${\rm SNR}=879$, while the
system at redshift $z=3$ had $m_1= 2 \times 10^6 M_{\odot}$, $m_2= 3 \times 10^6 M_{\odot}$ ($\eta=0.24$) and
${\rm SNR}=280$.

Figures~\ref{alpha-bounds}-\ref{beta-bounds} are `exclusion' plots, showing the 
region (above the curves) which could be excluded with a $99.73 \%$ confidence.
These figures also plot the bound on the ppE parameters that have already been achieved through analysis of the
`golden pulsar' system, PSR J0737-3039 ~\cite{2010PhRvD82h2002Y}. Observe that for the amplitude parameter $\alpha$, the
pulsar bounds beat the aLIGO bounds through almost the entire range of $a$; LISA can improve
upon the pulsar bounds for $a>0$. For the phase parameter $\beta$, however, both aLIGO and LISA
do better than the pulsar analysis through a significant portion of the range. As expected, gravitational wave observations tend
to do better in the strong field regime, corresponding to high post-Newtonian terms ($b > -5/3$ and $a > 0$), while the reverse is true for binary pulsar observations.

Vertical lines in Figs.~\ref{alpha-bounds} and \ref{beta-bounds} can be mapped to bounds on specific 
alternative theories, which we can then compare to current Solar System constraints. For example, consider
the following cases:
\begin{itemize}
\item Brans-Dicke [$(\alpha, b, \beta_{\BD}) = (0, -7/3, \beta_{\BD})$]: The tracking of the Cassini spacecraft~\cite{Bertotti:2003rm} has constrained $\omega_{\BD} > \bar{\omega}_{\BD} \equiv 4 \times 10^{3}$, which then forces $\beta_{\BD} < (5/3584) 4^{-2/5} (s_{1} - s_{2})^{2}/\bar{\omega}_{\BD}$, where $s_{1,2}$ are the sensitivities of the binary components (for BHs $s_{\BH} = 1/2$, and for NSs $s_{\NS} \approx 0.2-0.3$). 
\item Massive Graviton [$(\alpha, b, \beta_{\MG}) = (0, -1, \beta_{\MG})$]: Observations of Solar system dynamics~\cite{Talmadge:1988qz} have constrained $\lambda_{\MG} > \bar{\lambda}_{\MG} \equiv 2.8 \times 10^{12} \; {\rm{km}}$, which then forces $\beta_{\MG} < \pi^{2} (D/\bar{\lambda}_{\MG}) {\cal{M}} (1 + z)^{-1} \; {\rm{km}}^{-2}$, where $D$ is a distance measure to the source~\cite{Will:1997bb}.
\end{itemize}
The Solar System constraint on $\beta_{MG}$ is shown in Fig.~\ref{beta-bounds} 
with a black circle\footnote{We don't show similar constraints 
for Brans-Dicke theory, as here we consider binary BH inspirals, for which the Brans-Dicke 
correction would vanish due to the no-hair theorem.}. Observe that the constraints we could place with aLIGO and particularly LISA can be
orders of magnitude stronger than Solar System constraints (below the black circle). This is more easily seen by mapping our projected
constraints on $\beta_{\MG}$ to constraints on $\lambda_{\MG}$; with the aLIGO source, 
we find $\lambda_{\MG} \lesssim 8.8 \times 10^{12} \; {\rm{km}}$, while for the LISA source,
we find $\lambda_{\MG} \lesssim 3.763 \times 10^{16} \; {\rm{km}}$. This is consistent with results from previous
Fisher~\cite{Will:1997bb,Will:2004xi,Berti:2005qd,Stavridis:2009mb,Arun:2009pq,Keppel:2010qu,Yagi:2009zm,Will:1994fb,Scharre:2001hn} 
and Bayesian studies~\cite{2011arXiv1101.1391D}. Plotted for comparison are the bounds from  Pozzo et al. ~\cite{2011arXiv1101.1391D} on the upper panel of ~\ref{beta-bounds} and from Stavridis and Will ~\cite{Stavridis:2009mb} on the lower panel of~\ref{beta-bounds} both labeled as ``massive graviton." We find that our bound on $\beta$ for $b = -1$ is quite comparable to those found in these previous studies. Finally, shown on the lower panel of Fig.~\ref{beta-bounds} are the bounds found in the study by Arun et~al.~\cite{Arun:2009pq}, which allowed the PN coefficients themselves to vary as parameters. Their bounds on $\beta$ are somewhat weaker than those we found in our analysis, but this is an expected effect of the covariance between the PN coefficients. 

For all comparisons with previous studies, we took into account differences in SNR between the systems we analyzed and those we were comparing to. We also chose systems with the same or very similar total masses and mass ratios as those explored in previous papers. For the LISA systems, we compare the results from previous papers to our results for redshift $z=1$.

These plots show several other features that deserve further discussion. First, observe that all results show very 
little dependence on the choice of system parameters. This is quantitatively true for the aLIGO sources, shown in the upper panels of  Figs.~\ref{beta-bounds} 
and~\ref{alpha-bounds}, as these signals have the same SNR.  The LISA sources, shown in the lower panels of Figures~\ref{beta-bounds} 
and~\ref{alpha-bounds}, show a factor of $\sim 9$ offset, since these curves correspond to signals with different
SNRs. The SNR difference is a factor of $\sim 3$, which is a bit surprising as one would expect 
the spread on a parameter to scale with the SNR, and not the square of the SNR. However, we are working 
here in a region where the quadratic approximation to the Fisher matrix is singular, so the usual scaling does not hold.
The more rigorous bounds derived in the next section do follow a linear scaling with SNR, which
is reasonable since they use ppE injections and have non-singular Fisher matrix elements for the
ppE parameters.

Another interesting feature in these plots are the spikes at certain values of $a$ and $b$. These spikes say that for those
values of $a$ and $b$, gravitational wave observations can say little about the magnitude of GR deviations. The reason
for such spikes is that for those values of $a$ and $b$, $\alpha$ and $\beta$ become completely or partially degenerate 
with other parameters. For instance, when $a = 0$, $\alpha$ is fully
degenerate with the luminosity distance, and when $b=0$, $\beta$ is fully degenerate with the
initial orbital phase $\phi_c$. 

\begin{figure}[h]
\includegraphics[width=88mm,clip=true,angle=0]{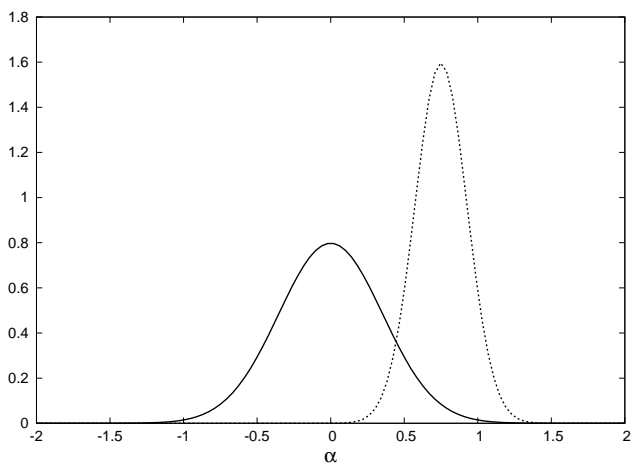}
\caption{\label{bound-comparison}An illustration of the two approaches for calculating cheap bounds on the ppE amplitude
parameters. The solid curve illustrates the bound that can be derived
by looking at the spread in the amplitude $\alpha$ when applying the ppE search to GR signals. In this
example, values of $\vert \alpha \vert > 1.5$ would be taken as indicating a departure from GR. The
dashed curve shows the bound that can be derived by starting with ppE signals and
determining how large the ppE amplitude needs to be for the posterior
distribution to have little weight at the GR value of $\alpha=0$. In this example, theories with
$\alpha > 0.75$ would be considered distinguishable from GR.}
\end{figure}

One can also develop `cheap' bounds that use ppE instead of GR signal injections. For instance, one could start with
injections with a range of values for $\alpha$ and $\beta$, and then look to see when
the posterior distributions for these parameters no longer show significant support
at the GR values of $\alpha=\beta=0$. These two types of cheap bounds are
illustrated in Fig.~\ref{bound-comparison}. Given an observation of a non-zero $\alpha$, a cheap bound
calculation as described in this section (solid curve) would indicate
a value $\vert \alpha \vert  < 1.5$ is still
consistent with GR. A similar study with ppE injections, however, which produced the dashed-curve posterior distribution for $\alpha$ would indicate a preference for the ppE model over the GR model with a detection of  $\alpha > 0.75$. Thus the technique used in this section, which is a variance-covariance study, answers an inherently different question from a model selection study. In the next section, we explore model selection in detail.

\subsection{Rigorous Bounds and Model Selection}

In order to see how accurate the cheap bounds found in the previous section are, we next performed
a full Bayesian model selection analysis on several different signals. We injected a signal with a
given set of ppE parameters and ran a search using both GR and ppE templates. We then calculated
the Bayesian evidence for each model and from this the Bayes factor. To compare these results to
the cheap bounds, we ran the analysis on several different ppE signals, each with the same injected
value of $a$ or $b$, but with progressively larger values of $\alpha$ or $\beta$. This then allows
us to determine the values of ppE amplitudes $\alpha$ or $\beta$ where the evidence for the ppE
hypothesis exceeds that of the GR hypothesis by some large factor, which we took to be Bayes factors
in excess of 100 (in the Jeffery's classification~\cite{1951ZNatA...6..471J}, Bayes factors in
excess of $100$ represent {\emph{decisive}} evidence in favor of that model).

We do not expect the cheap bounds to agree precisely with the more rigorous model selection bounds
as they are based on quite different reasoning. The cheap bounds simulate what we would find if GR
was consistent with observations, and establishes the spread in the ppE amplitude parameters that would
remain consistent. If we were to analyze some data and find ppE amplitude parameters outside of
this range, it would give us motivation to search more rigorously for departures from GR. With the more
expensive model selection bounds, we start with non-GR signals and seek to determine how large the departures
from GR have to be for the ppE hypothesis to be preferred. In the first case the distribution of $\alpha$ and
$\beta$ is known to be centered around zero, but in the second case they are not, so the two analyses
should not be expected to agree precisely.

One can derive a more detailed connection between the alternative form of
the cheap bounds derived using ppE injections (discussed at the end of the previous section) and the more
rigorous Bayesian evidence calculations
using the Savage-Dickey density ratio~\cite{1995}. The latter states that for nested hypotheses with separable priors,
the Bayes factor is equal to the ratio of the posterior and prior densities evaluated at the parameter
values that correspond to the lower dimensional model. If the posterior distribution was a Gaussian with
width $\sigma$ centered at $\alpha= n \sigma$, and we were using a uniform prior with width $N\sigma$, then
the Bayes factor would equal ${\rm{BF}}=Ne^{-n^2/2}/\sqrt{2\pi}$, where this Bayes factor shows the odds of the lower
dimensional model being correct. For example, with $N=100$ and $n=4$ we get a
Bayes factor of ${\rm{BF}}=0.013$, showing strong support for the higher dimensional model. While the
cheap bounds that can be derived using ppE signal injections will be stronger than the cheap bounds
that can be derived from GR signal injections, the computational cost is higher as multiple simulations
have to be run to find the transition point, and this approach is only moderately cheaper than
performing the full Bayesian model selection.

\begin{figure}[h]
\includegraphics[width=88mm,clip=true,angle=0]{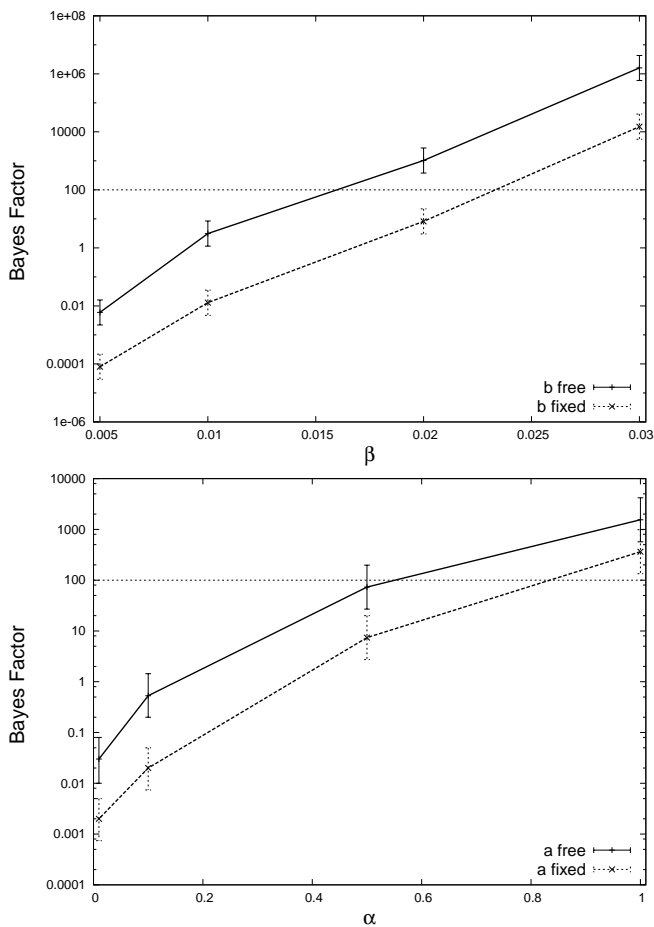}
\caption{\label{BF-aLIGO} UPPER PANEL: Bayes factors for a ${\rm SNR}=20$ aLIGO ppE injection with parameters $(a, \alpha, b, \beta) = (0, 0, -1.25, \beta)$. The Bayes factors are the 'betting odds' that ppE  (and not GR) is the model that accurately describes the data. As the deviation from GR gets larger, ppE becomes the preferred model. \newline
LOWER PANEL: Bayes factors for a ${\rm SNR}=20$ aLIGO ppE injection with parameters $(a, \alpha, b, \beta) = (-0.5, \alpha, 0, 0)$.}
\end{figure}

Examples of the full model selection procedure are shown in Fig.~\ref{BF-aLIGO} for aLIGO/aVirgo
detections with ${\rm SNR} = 20$. Each panel shows Bayes factors for two types of ppE search,
one with $a$ or $b$ held fixed at the injected value, and one in which all four ppE values were allowed
to vary. The Bayes factor, defined in Eq.~\eqref{BF-def}, is here the odds ratio between the ppE model and the GR model. A larger Bayes factor indicates a stronger preference for the ppE model. The search in which $a$ or $b$ was fixed provides the closest comparison with the cheap
bounds of the previous section. The bound on $\beta$ derived by setting a Bayes factor threshold
of 100 are roughly 3 times larger than the cheap bounds when $b$ is held fixed and roughly 2 
times larger when $b$ is free to vary. The bounds on $\alpha$ match the cheap bounds when $a$ is held
fixed, and is slightly smaller when $a$ is allowed to vary. 

\begin{figure}[h!]
\includegraphics[width=88mm,clip=true,angle=0]{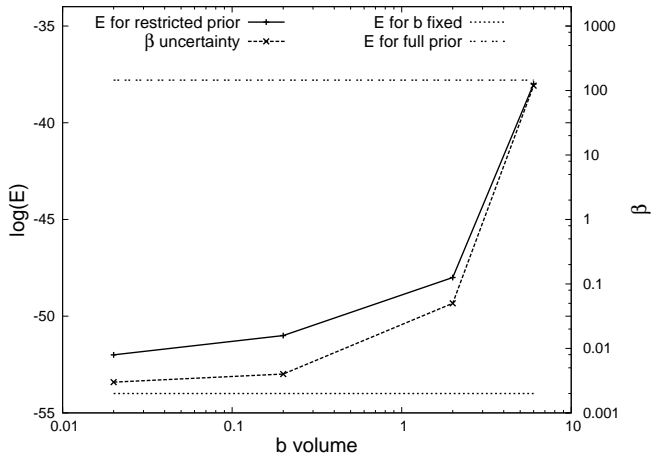}
\caption{\label{interpolation-plot} Here we plot the log of the evidence (E) for the ppE model characterizing a ppE injection as the prior volume on $b$ is increased. The evidence for the ppE model increases with the prior volume on $b$. The growth
in the evidence can be attributed to the growth in the variance of $\beta$, which lessens the severity of the `Occam penalty' for more model parameters.}
\end{figure}

We were surprised to find that the bounds are
tighter for the higher dimensional models, with ($a,b$) free, than for the lower dimensional models, with
($a,b$) fixed. To explore this more thoroughly, we performed a study where the prior on $b$ was increased
from a very small range to the full prior range. Since holding a parameter fixed is equivalent to
using a delta-function prior, we expect the evidence to interpolate between the values found when $b$ was fixed and when $b$ was free to explore the full prior. Figure~\ref{interpolation-plot} confirms this expectation, and also provides an explanation for the
growth in the evidence. 

To understand this plot, it is helpful to look at the Laplace approximation to
the evidence~\cite{azevedo-filho:laplace's}, which assumes that the region surrounding the
maximum of the posterior distribution
is well approximated by a multivariate Gaussian. With this assumption, the evidence is given by
\begin{equation}\label{laplace}
p(d| \mathcal{H}) \approx p(d\vert \vec{\theta}, \mathcal{H})\vert_{\rm MAP}
\, \left(\frac{\Delta V_\mathcal{H}}{V_\mathcal{H}}\right) \, .
\end{equation}
The first term is the likelihood evaluated at the maximum of the posterior, and the
second term is the ratio of the posterior volume $\Delta V$ to the prior volume $V$.
The posterior volume can be estimated from the volume of the error ellipsoid
containing 95\% of the posterior probability. The ratio ${\cal O}=\Delta V/V$ is termed the ``Occam factor'',
and the quantity $I = \log_2 (V/\Delta V)$ provides a measure of how much information has been gained
about the parameters from the data. 

Now consider a situation where we have nested hypotheses
${\cal H}_0$ and ${\cal H}_1$, with the second hypothesis involving an additional parameter $y$.
If the likelihood is insensitive to $y$ then the first factor in the evidence stays the same, and
since $y$ is unconstrained, $\Delta V_y = V_y$ and the Occam factor is also unchanged. Thus,
both models have the same evidence, even though one has more parameters than the other. Conversely,
if the additional parameter is tightly constrained by the data, $\frac{\Delta V_y} {V_y}$ can be a very small
number. In this case, the evidence for ${\cal H}_1$ is much reduced by the Occam factor, and the factor is
referred to as an ``Occam penalty."

The growth in evidence for the ppE model as the prior range for $b$ gets larger is an effect of this Occam factor, which is a ratio
of the uncertainty in the recovered value of an extra parameter to the prior volume for that parameter. As the prior range
on $b$ expands, this leads to a greater variance in the recovered values for $\beta '$.  Because the prior volume of $\beta '$ remains unchanged,
the large growth in its variance as the prior range of $b$ is expanded leads to a large growth in the Occam factor - and thus a shrinking of the Occam penalty. As the
Occam factor gets larger, so does the evidence for the ppE model.
The evidence for the GR model, of course, does not depend
on the priors we use for the ppE parameters, and so as the evidence for ppE grows, the Bayes factor indicates a stronger
preference for ppE.

\begin{figure}[h!]
\includegraphics[width=88mm,clip=true,angle=0]{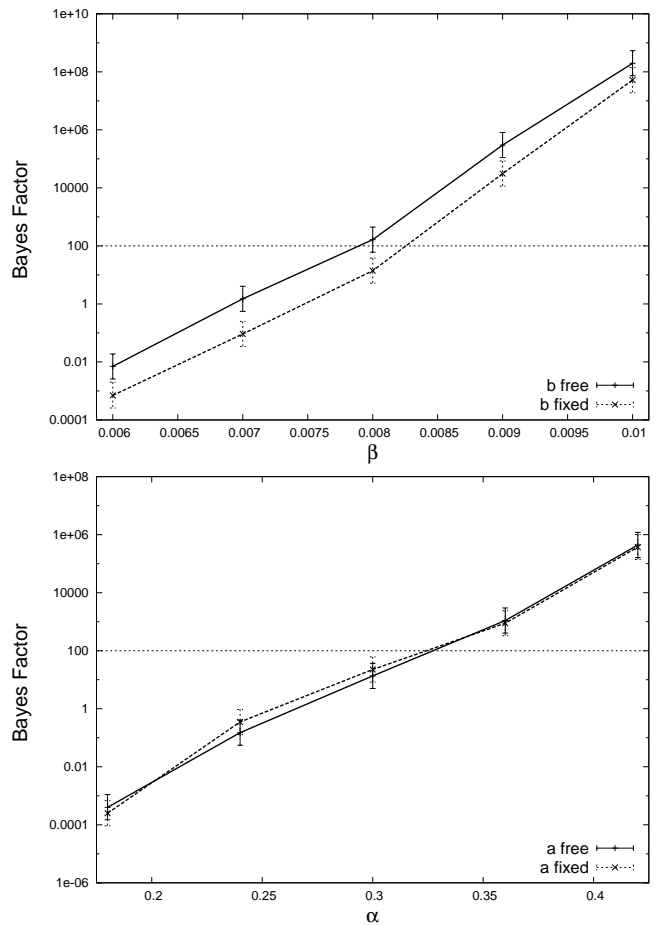}
\caption{\label{BF-LISA} UPPER PANEL: Bayes factors for a $z=1$ LISA ppE injection with parameters $(a, \alpha, b, \beta) 
= (0, 0, -1.0, \beta)$. \newline
LOWER PANEL: Bayes factors for a $z=1$ LISA ppE injection with parameters $(a, \alpha, b, \beta) 
= (0.5, \alpha, 0, 0)$.}
\end{figure}

Figure~\ref{BF-LISA} shows Bayes factors between the GR and ppE hypotheses for a $z=1$ LISA source.
In the upper panel, the injections where chosen with $a=0, b=-1$ and variable $\beta$, while in the lower panel the
injections were chosen with $a=0.5, b=0$ and variable $\alpha$. Because LISA sources have much
higher SNR, the ppE parameters are more tightly constrained, and the difference between the Bayes
factors when $a$ or $b$ are fixed versus freely varying is less pronounced. The more rigorous bounds on
$\alpha$ and $\beta$ are both a factor of $\sim 2$ times weaker than those predicted by the cheap bounds, which
is in line with what we found for the phase correction $\beta$ in the aLIGO example. In summary, the
cheap bounds provide a fair approximation to the bounds that can be derived from Bayesian model selection,
and can generally be trusted to within an order of magnitude.

\subsection{Fitting Factor}

Another quantity of interest is the fitting factor, which measures how well one template family can
recover an alternative template family. To define the fitting factor, we must first define the match between
two templates $h$ and $h'$ as
\begin{equation} \label{match}
{\cal M} = \frac{(h|h')}{\sqrt{(h|h)}\sqrt{(h'|h')}} \, .
\end{equation}
The match is related to the metric distance between templates~\cite{1996PhRvD53.6749O}
by ${\cal M}=1-\frac{1}{2} g_{ij}\Delta x^i \Delta x^j$, where the metric is evaluated
with the higher-dimensional model (appropriate when dealing with nested models).
The fitting factor ${\rm FF}$ is then defined as the best match that can be achieved by varying the
parameters of the $h'$ template family to match the template belonging to the
the other family, $h$. 

Another interpretation for the fitting factor is as the fraction
of the true signal-to-noise ratio ${\rm SNR} = \sqrt{(h|h)}$ that is recovered
by the frequentist statistic $\rho = {\rm max}[(h|h')/\sqrt{(h'|h')}]$.
The imperfect fit leaves behind a residual $(h-h')$ with ${\rm SNR}^2_{\rm res}=\chi^2$,
which can be minimized by adjusting the amplitude of $h'$ to yield
\begin{equation}
{\rm SNR}_{\rm res}^2= (1-{\rm FF}^2) {\rm SNR}^2 \, .
\end{equation}
Assuming that a residual with ${\rm SNR}_*$ is detectable, and working in the limit where ${\rm FF} \sim 1$, 
we have
\begin{equation}
1-{\rm FF} \simeq \frac{{\rm SNR}^2_{*}}{2 \, {\rm SNR}^2} \, .
\label{FF-eq}
\end{equation}

\begin{figure}[h]
\includegraphics[width=88mm,clip=true,angle=0]{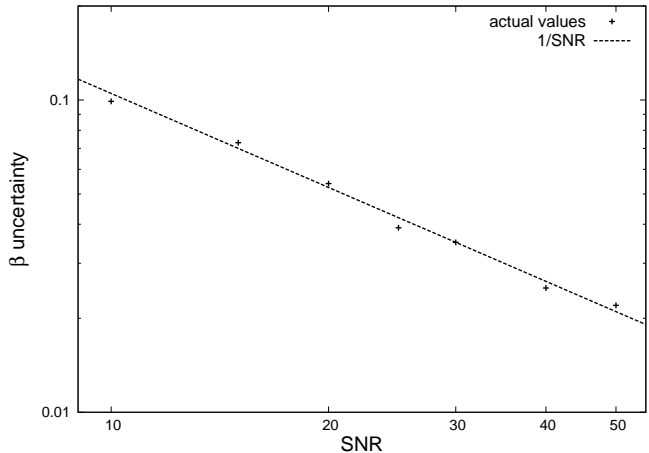}
\caption{\label{SNRstudy} The scaling of the parameter estimation error in the ppE parameter $\beta$ for an aLIGO
simulation with ppE parameters $(a, \alpha, b, \beta) = (0, 0, -1.25, 0.1)$. The parameter errors follow
the usual $1/{\rm SNR}$ scaling.}
\end{figure}

We see then that the ability to detect departures from GR scales inversely with the square of the SNR, 
as given by Eq.~\eqref{FF-eq}. On the other hand, the detectable difference between the parameters 
in the two theories will scale inversely with a single power of the SNR. This is because this detectable 
difference is proportional to the square-root of the minimized match function and  
 \begin{equation}
\sqrt{{\rm min}(g_{ij}\Delta x^i \Delta x^j)} \simeq \frac{{\rm SNR}_{*}}{{\rm SNR}} \, ,
\end{equation}
and the metric is independent of SNR. This reasoning applies to both the additional model parameters
of the alternative theory, {\it e.g.} $\Delta x^{i} = (\alpha,\beta)$, and the physical source parameters
such as the masses and distance. We then expect both the bounds on the ppE model parameters
and the biases caused by using the wrong template family to scale inversely with SNR. This scaling is
in keeping with the usual scaling of parameter estimation errors that follows from a Fisher matrix
analysis where $\langle \Delta x^i \Delta x^j \rangle \simeq (h_{,i} \vert h_{, j})^{-1} \sim {\rm SNR}^{-2}$.
Figure~\ref{SNRstudy} shows that the errors in the recovery of the ppE parameters follows the expected
scaling with SNR.

\begin{figure}[h]
\includegraphics[width=90mm,clip=true,angle=0]{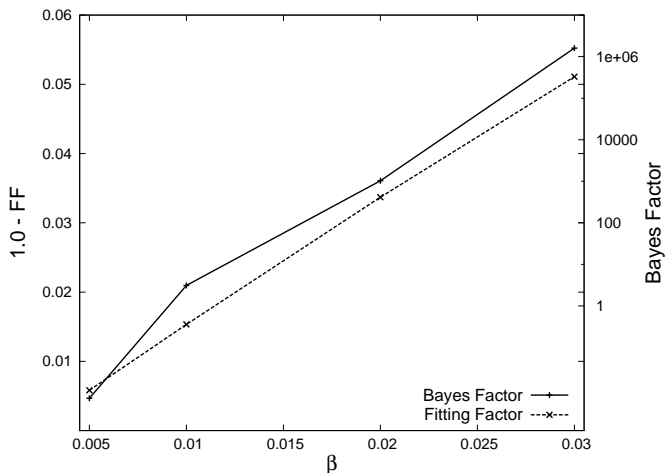}
\caption{\label{bayesstudy} The log Bayes factors and $(1 - {\rm FF})$ plotted as a function of $\beta$ for a ppE injection with
parameters $(a, \alpha, b, \beta) = (0, 0, -1.25, \beta)$.  The predicted link between the fitting
factor and Bayes factor is clearly apparent.}
\end{figure}

Alternative models that are not well-fitted by GR will be more easily distinguished than
models that can be well-fitted. This suggests that there should be a correlation between the
fitting factor and the Bayes factor. The relationship can be established using the
Laplace approximation to the evidence [Eq.~(\ref{laplace})], from which it follows that the log Bayes factor
is equal to
\begin{eqnarray}
\log {\rm{BF}} & = &\log{ \frac{e^{-\chi ^2 (\mathcal{H} _1)/2}}{e^{-\chi ^2 (\mathcal{H}_0)/2}} \frac{\mathcal{O} _1}{\mathcal{O}_0}} \nonumber \\
	& = & \frac{\chi^2_{\rm min}}{2} + \Delta \log {\cal O}  \nonumber \\
       & = & (1-{\rm FF}^2) \frac{{\rm SNR}^2}{2} + \Delta \log {\cal O} \,,
\end{eqnarray}
where ${\cal{O}}$ is the Occam factor, defined in the discussion following [Eq.~(\ref{laplace})]. Thus, up to the difference in the log Occam factors, 
the log Bayes factor should scale as $2(1-{\rm FF})$ when ${\rm FF} \sim 1$. This link is confirmed in Figure~\ref{bayesstudy}.

\subsection{Parameter Biases}

\begin{figure}[h!!]
\includegraphics[width=90mm,clip=true]{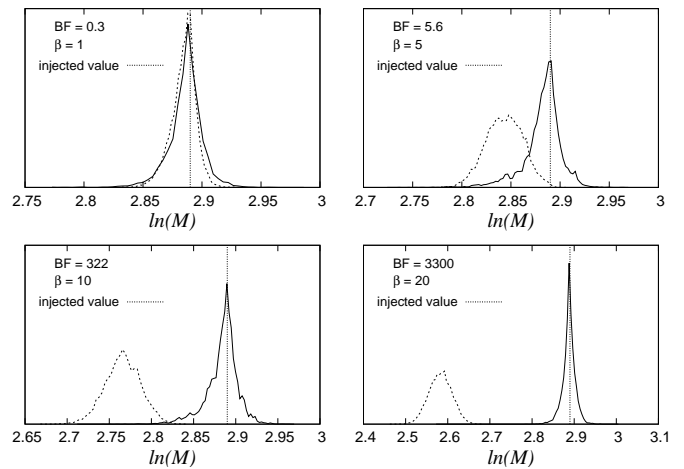}
\caption{\label{total-mass-bias} Histograms showing the recovered log total mass for GR (dashed) and
ppE (solid) searches on ppE signals. As
the source gets further from GR, the value for total mass recovered by the GR search moves away from
the actual value. All signals had injected $b = -0.25$.}
\end{figure}

If we assume that Nature is described by GR, but in truth another theory is correct, this will
result in the recovery of the wrong parameters for the systems we are studying. For instance, when looking
at a signal that has non-zero ppE 
phase parameters, a search using GR templates will return the incorrect mass parameters, as illustrated in
Fig.~\ref{total-mass-bias}. Observe that as the magnitude of $\beta$ is increased (thus increasing the Bayes factor), 
the error in the chirp mass parameter extraction grows well beyond statistical errors. 

Perhaps the most interesting point to be made with this study is that the GR templates return values of
the total mass  that are completely outside the error range of the (correct) parameters returned by
the ppE search, \emph{even for ppE signals that are not clearly discernible from GR}. We refer to this parameter
biasing as `stealth bias', as it is not an effect that would be easy to detect, even if one were looking for it. 

\begin{figure}[h!!]
\includegraphics[width=117mm,clip=true]{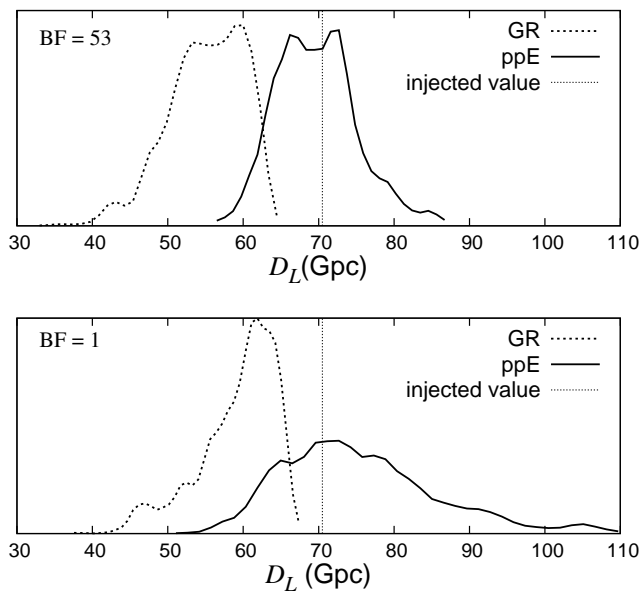}
\caption{\label{DL-bias} Histograms showing the recovered values for luminosity distance from GR and ppE searches on a LISA
binary at redshift $z=7$. Both signals have $a=0.5$, and were injected with a luminosity distance of 70.5 Gpc.
The top plot has $\alpha = 3.0$ and the bottom has $\alpha = 2.5$.
As the Bayes factor favors the ppE model more strongly, the bias in the recovered luminosity distance
from the GR search becomes more pronounced.}
\end{figure}

As an example, consider stealth bias for non-zero ppE $\alpha$ parameters, as illustrated in Fig.~\ref{DL-bias}. 
As one would expect, when a GR template is used to search on a ppE signal that has non-zero ppE amplitude corrections, 
the parameter that is most affected is the luminosity distance. We again see the bias of the recovered parameter becoming more
apparent as the signal differs more from GR\footnote{Here, the uncertainty in the recovered luminosity distance
changes considerably between the different systems, because we held the injected luminosity distance
constant instead of the injected SNR.}. For example, the recovered posterior
distribution from the search using GR templates has zero weight at the correct value of luminosity distance when 
the Bayes factor is $\sim 50$. Even when the Bayes factor is of order unity, the peaks of the posterior distributions of 
the luminosity distance differ by approximately $10$ Gpc.

\section{Conclusion}
\label{sec:Conclusions}

The two main results of this study are that GW observations of binary compact object inspirals using ppE
waveforms can constrain higher PN order (i.e. $b>-5/3$ and $a>0$) deviations from GR much more
tightly than binary pulsar observations,
and that parameter estimates can be significantly biased if GR templates are used
to recover signals when an alternative theory of gravity better describes the event.
This latter bias can be significant even in cases where it is {\em not}
obvious that GR is not quite the correct theory of gravity. We also see that the detection efficiency of GR templates
can be seriously compromised if they are used to characterize data that is not described by GR.

The current study makes several simplifying assumptions about the waveforms: we
consider only the inspiral stage for non-spinning black holes on circular orbits, and 
include just the leading order ppE corrections to the waveforms. In future work we plan
to include a marginalization over these higher order corrections. Including this marginalization
will be more realistic, as the ppE formalism allows for many higher order corrections to the waveform.
Marginalizing over the higher order terms will weaken the bounds on the leading order ppE parameters,
though probably not by that much since they are sub-dominant terms.

Another subject that we will examine in the future is the effect on our analysis of multiple detections.
Simultaneously characterizing several systems with different mass ratios should allow us to examine the dependence
of the $\alpha$/$\beta$ parameters on spin, mass difference, mass ratio, etc.. Furthermore, looking at several
systems simultaneously will break the degeneracies between the ppE parameters and the individual system parameters
(masses, distances {\it etc}), and will allow us to detect significantly smaller deviations from GR.

We also plan to perform a study similar to that done by
Arun {\it et al.}~\cite{Arun:2006hn,Arun:2006yw,Mishra:2010tp}, in which
the exponents $a_i,b_i$ are fixed at the values found in the PN expansion of GR, and compare their
Fisher matrix based bounds to those from Bayesian inference. We expect a full Bayesian inference study
to lead to significantly different conclusions, due to the singularities in the Fisher matrix already observed in 
the present study. 

Finally, we will look at LISA observations of galactic white-dwarf binaries to see if the brighter
systems, which may have SNRs in the hundreds, may allow us to beat the pulsar bounds across the
entire ppE parameter space. The brightest white-dwarf systems will have $u \sim 10^{-8} \rightarrow 10^{-7}$
(for comparison the `golden' double pulsar system, PSR J0737-3039A has $u = 3.94 \times 10^{-9})$, and
these small values for $u$ make the ppE effects, which scale as $u^a$ and $u^b$, much larger than
for black hole inspirals when $a,b < 0$.

The chance to test the validity of Einstein's theory of gravity is one of the most exciting opportunities that 
gravitational wave astronomy will afford to the scientific community. Without the appropriate tools, however,
our ability to perform these tests is sharply curtailed. This analysis has shown that the ppE template family
could be an effective means of detecting and characterizing deviations from GR, and also that assuming
that our GR waveforms are correct could lead to lessened detection efficiency and biased parameter estimates
if gravity is described by an alternative theory (even when choosing parameters at the threshold of what has
already been ruled out by Solar System and binary pulsar observations).
We have identified several areas of future investigation, and will continue to study this area in depth.

\acknowledgments

We thank Patrick Brady, Curt Cutler, Ben Owen, David Spergel, Xavier Siemens, 
Paul Steinhardt and Michelle Vallisneri for detailed comments and suggestions.
We are very grateful to Martin Weinberg and Will Farr for making their direct
evidence integration codes available to us, and for helping us to understand the
results. N.~J. and L.~S.~acknowledge support from the NSF Award 0855407 and NASA grant
NNX10AH15G. N.~Y.~and F.~P.~acknowledge support from the NSF grant PHY-0745779, 
and FP acknowledges the support of the Alfred P. Sloan Foundation.

\bibliography{phyjabb,master}

\section{Appendix A}
As described in Section III, the VTA method for calculating evidences involves two possible sources of
error. One is introduced by the fact that our Markov chains are of finite length. To get an idea of the
magnitude of this statistical uncertainty, the implementation of the VTA that we used calculates the evidence
many times using different sub-samples of the Markov chain. This process is called bootstrapping, and we find
that in general it results in an uncertainty in the log Bayes factor of the order $\pm 0.5$.  

The second source of possible error in the VTA techniques comes from the choice of boxing number. The boxing
number is the number of points from the chain that are sorted into each volume element. A higher boxing number
will return a more accurate number for the mean or median of the posterior in a given volume element, but at
the cost of having large volume elements that may not resolve fine features in the posterior distribution.
Lower boxing numbers lead to greater variance in the estimate of the posterior density in each cell, but
allows for better resolution of sharp features in the posterior landscape. To examine the systematic
error in Bayes factors associated with using different boxing numbers, we calculated the Bayes factor
between ppE and GR models for a source with injected ppE parameters $(a,\alpha,b,\beta) = (0.5,75,0,0)$.
We first used thermodynamic integration with a run using 50 chains, and found the log Bayes factor to
equal $\log(B) = 12.0 \pm 1.0$. Because thermodynamic integration performs more accurately than the VTA
when integrating posterior distributions for which analytic answers are available, such as a multi-variate
Gaussian, we take this value as our reference. We then calculated the log Bayes factor using the VTA with
boxing numbers of 16, 32, and 64. The results, including the statistical uncertainty, are shown in Table II.

\begin{table}[h!]\label{tab2}
\caption{Bayes factors calculated using the VTA with different boxing numbers.
} 
\label{aggi}\centering
\begin{tabular}{|c|c|c|c|}\hline
Boxing Number        & $\; GR \;$				& $\; ppE\;$			     	& $\; log(BF)\;$   	\\ \hline
16                                & $-41.50^{+0.2}_{-0.22}$		& $-31.04^{+0.88}_{-0.75}$       & 	 $10.5^{+1.0}_{-1.0}$          			\\
32                                & $-40.43^{+0.15}_{-0.13}$	&$-28.02^{+0.67}_{-0.44} $     	&   $12.4^{+0.8}_{-0.6}$	                  		 \\
64                                &$ -39.51^{+0.26}_{-0.31}$	& $-25.78^{+0.43}_{-0.30} $      &   $13.7^{+0.7}_{-0.6} $       			\\
\hline
\end{tabular}
\end{table}

The results show that the variation in $\log(B)$ between different boxing sizes is similar to, but
slightly larger than the statistical variation introduced by the VTA within one boxing size. The
variation due to choice of boxing size is roughly $\pm 1.5$. We therefore use error bars indicating
$\log(B) \pm 1$ on our Bayes factor plots. Further, we found that a boxing size of 32 returned the
most accurate value for the Bayes factor, and so we used this size for the rest of our analysis.

\end{document}